\documentclass[10pt,lettersize,journal]{IEEEtran}
\usepackage{amsmath,amsfonts}
\usepackage{algorithm}
\usepackage{array}
\usepackage[caption=false,font=normalsize,labelfont=sf,textfont=sf]{subfig}
\usepackage{textcomp}
\usepackage{stfloats}
\usepackage{url}
\usepackage{verbatim}
\usepackage{graphicx}
\usepackage{cite}
\usepackage{amsmath}
\usepackage{algpseudocode}
\usepackage{tikz}
\usepackage{verbatimbox} 
\usepackage{graphicx}
\usepackage{textcomp}
\usepackage{xcolor}
\usepackage{listings}
\usepackage{multirow}
\usepackage{caption}
\usepackage{pifont}
\usepackage{graphicx}
\usepackage{booktabs}
\usepackage{colortbl}
\usepackage{array}
\usepackage{multirow}
\usepackage{hyperref}
\usepackage{makecell}
\usetikzlibrary{positioning, arrows.meta, shapes.geometric}

\usepackage[a4paper, total={184mm,239mm}]{geometry}
\def\BibTeX{{\rm B\kern-.05em{\sc i\kern-.025em b}\kern-.08em
    T\kern-.1667em\lower.7ex\hbox{E}\kern-.125emX}}
    
\definecolor{keywordcolor}{rgb}{0, 0, 0}
\definecolor{commentcolor}{rgb}{0.5, 0.5, 0.5}
\definecolor{stringcolor}{rgb}{0.5, 0.5, 0.5}
\definecolor{backgroundcolor}{rgb}{0.97, 0.97, 0.97} 
\lstdefinestyle{customverilog}{
  language=Verilog,
  basicstyle=\ttfamily\footnotesize,
  keywordstyle=\color{keywordcolor}\bfseries,
  commentstyle=\color{commentcolor},
  stringstyle=\color{stringcolor},
  morekeywords={module, if, else, case, end, begin, endcase, input},
  showstringspaces=false,
  numberstyle=\tiny\color{red},
  frame=single,
  rulecolor=\color{black},
  backgroundcolor=\color{white},
  tabsize=2,
  captionpos=b,
  breaklines=true,
  postbreak=\mbox{\textcolor{red}{$\hookrightarrow$}\space},
}

\usepackage[utf8]{inputenc}
\usepackage{xcolor}   
\usepackage{listings} 

\definecolor{codegray}{rgb}{0.5,0.5,0.5}

\usepackage[most]{tcolorbox}
\usepackage{listings}
\usepackage{xcolor}
\usepackage{enumitem}

\newtcolorbox{verificationbox}[2][]{
    enhanced,
    colframe=#2!75!black,
    colback=#2!3!white,
    arc=2mm,
    boxrule=0.8pt,
    left=3mm, right=3mm, top=2mm, bottom=2mm,
    drop shadow={shadow xshift=0.3mm, shadow yshift=-0.3mm, opacity=0.2},
    title={#1}, 
    fonttitle=\bfseries\small,
    coltitle=#2!90!black,
    attach boxed title to top left={xshift=0.5cm, yshift=-2mm},
    boxed title style={
        size=small,
        colback=#2!20!white,
        colframe=#2!70!black,
        arc=1.5mm
    },
    overlay first={
        \node[#2!75!black] at ([xshift=5mm]frame.north west) 
            {\scriptsize\faIcon[solid]{code}};
    }
}

\begin{document}

\title{AutoVeriFix+: High-Correctness RTL Generation via Trace-Aware Causal Fix and Semantic Redundancy Pruning}
\author{Yan~Tan,~\IEEEmembership{Student Member,~IEEE,}
        Xiangchen~Meng,~\IEEEmembership{Student Member,~IEEE,}\\
        Zijun~Jiang,~\IEEEmembership{Student Member,~IEEE,} 
        and~Yangdi~Lyu,~\IEEEmembership{Member,~IEEE}
\thanks{This work was supported by the National Natural Science Foundation of China under Grant \#62402412. \em{(Corresponding author: Yangdi Lyu.)}}
\thanks{Yan Tan, Xiangchen Meng, Zijun Jiang, and Yangdi Lyu are with the Microelectronics Thrust, the Hong Kong University of Science and Technology (Guangzhou), Guangzhou 511453, China (email: ytan910@connect.hkust-gz.edu.cn; xmeng027@connect.hkust-gz.edu.cn; zjiang438@connect.hkust-gz.edu.cn; yangdilyu@hkust-gz.edu.cn;).}
}
\maketitle

\begin{abstract}
Large language models (LLMs) have demonstrated impressive capabilities in generating software code for high-level programming languages such as Python and C++. However, their application to hardware description languages, such as Verilog, is challenging due to the scarcity of high-quality training data. Current approaches to Verilog code generation using LLMs often focus on syntactic correctness, resulting in code with functional errors. To address these challenges, we propose AutoVeriFix+, a novel three-stage framework that integrates high-level semantic reasoning with state-space exploration to enhance functional correctness and design efficiency. In the first stage, an LLM is employed to generate high-level Python reference models that define the intended circuit behavior. In the second stage, another LLM generates initial Verilog RTL candidates and iteratively fixes syntactic errors. In the third stage, we introduce a Concolic testing engine to exercise deep sequential logic and identify corner-case vulnerabilities. With cycle-accurate execution traces and internal register snapshots, AutoVeriFix+ provides the LLM with the causal context necessary to resolve complex state-transition errors. Furthermore, it will generate a coverage report to identify functionally redundant branches, enabling the LLM to perform semantic pruning for area optimization. Experimental results demonstrate that AutoVeriFix+ achieves over 80\% functional correctness on rigorous benchmarks, reaching a pass@10 score of 90.2\% on the VerilogEval-machine dataset. In addition, it eliminates an average of 25\% redundant logic across benchmarks through trace-aware optimization.
\end{abstract}


\section{Introduction}
\label{subsex:intro}


\IEEEPARstart{L}{arge} language models (LLMs) have revolutionized software development by significantly enhancing productivity through automated code generation. By translating natural-language descriptions into high-quality software code, LLMs allow developers to focus on higher-level design and problem-solving rather than routine coding tasks. This productivity boost is largely attributed to the extensive training of LLMs on vast amounts of open-source repositories, particularly for widely used languages such as Python and C++~\cite{Mastropaolo_Pascarella_Guglielmi_Ciniselli_Scalabrino_Oliveto_Bavota_2023, Nijkamp_Hayashi_Xiong_Savarese_Zhou}. Consequently, LLMs currently achieve not only high syntactic accuracy but also robust functional correctness in the software domain.

Building on these successes, researchers have begun to explore the application of LLMs to generate Hardware Description Languages (HDLs). Initial efforts have shown promise in generating basic Register-Transfer Level (RTL) components~\cite{Benchmark_RTL_2022,Dehaerne_Verilog_2023,OpenLLM,VerilogEval,Xie_2023,2024origen}. However, a significant performance gap remains between software and hardware code generation~\cite{chen2021codex}. While HDLs share syntactic similarities with software languages, they require precise modeling of concurrent execution, timing constraints, and hardware-specific idioms. Furthermore, the relative scarcity of open-source Verilog datasets limits the efficacy of supervised fine-tuning for HDL tasks. Given that hardware bugs incur exponentially higher costs than software errors due to long Electronic Design Automation (EDA) tool runtimes and irreversible fabrication cycles, LLM-generated RTL must undergo rigorous functional verification before integration.

This performance discrepancy may be due to the fundamental differences in abstraction layers. Software languages such as Python operate at a level of functional abstraction that aligns well with the pattern-recognition capabilities of LLMs. In contrast, hardware description languages like Verilog require explicit specification of low-level circuit behaviors, particularly complex sequential logic and non-blocking assignments. To address this, recent studies have focused on hardware-specific data augmentation~\cite{Dehaerne_Verilog_2023,BetterV,Benchmark_RTL_2022,VerilogEval,OpenLLM}.
For instance, RTLCoder~\cite{Xie_2023} utilizes automated design-data augmentation to align natural language descriptions with Verilog semantics.
To further mitigate errors, self-reflection frameworks~\cite{AutoChip,2024origen,RTLFixer,verilogcoder} have been introduced to identify syntax errors and iteratively refine the generated code. 
However, existing methodologies primarily focus on syntactic corrections through pattern matching and template-based approaches, often failing to address deep functional discrepancies in multi-branch logic or complex Finite State Machines (FSMs).

To bridge this gap, we propose AutoVeriFix+, a three-stage framework that integrates an LLM-assisted automated testing mechanism with trace-aware optimization to ensure the functional integrity of RTL designs.
\begin{itemize}
    \item \textbf{Stage 1: Python-Assisted Reference Modeling.} We leverage the strong capabilities of existing LLMs in understanding and generating Python code to generate a high-level reference model from natural-language specifications. This model serves as a functional golden reference. While these models capture intended behavior, they cannot be directly synthesized into hardware due to limitations in the synthesizable subsets supported by High-Level Synthesis (HLS) tools such as MyHDL~\cite{myhdl} and SODA~\cite{soda}.
    \item \textbf{Stage 2: Iterative RTL Refinement.} In the second stage, a different LLM generates the initial Verilog RTL implementation, which then undergoes an automated syntax validation loop to resolve immediate compilation and linter errors. Once syntactically clean, the RTL is refined by integrating detailed debug information. Rather than relying on simple messages of I/O mismatch, we utilize causal feedback from the subsequent stage to specify precisely which internal logic is malfunctioning, ensuring the final code is both compilable and functionally aligned with the specification.
    \item \textbf{Stage 3: Trace-Aware Optimization and Pruning.} To move beyond limited black-box testing, we introduce a novel third stage that employs Concolic Testing to systematically explore deep logic corner cases and maximize branch coverage. By automatically instrumenting the RTL and parsing its Control Flow Graph (CFG), we align internal execution traces with the Python reference model. This white-box testing enables us to detect silent state-machine errors and eliminate redundant logic, thereby significantly improving the design's functional reliability and hardware efficiency.
\end{itemize}
 
Our main contributions are summarized as follows.
\begin{enumerate}
    \item We propose a framework that leverages the high-level reasoning of software-centric LLMs (Python) to serve as a functional oracle for the less mature hardware domain (Verilog), enabling automated golden-model generation.
    
    \item We introduce an auto-instrumentation mechanism that monitors runtime internal states. This white-box approach uncovers deep functional bugs that are invisible to traditional black-box testing.

    \item We develop a targeted feedback mechanism that provides LLMs with cycle-accurate traces and coverage reports, enabling precise bug fixing and the removal of redundant code.
    
    \item Our experimental results show that AutoVeriFix+ significantly outperforms state-of-the-art methods, achieving over 80\% functional correctness on rigorous benchmarks while simultaneously optimizing implementation efficiency.
    
\end{enumerate}

This paper is an extension of our preliminary study published in~\cite{autoverifix}. The current work, AutoVeriFix+, expands upon the original two-stage framework in several key dimensions. Specifically, we introduce a Stage 3 Trace-Aware Optimization and Pruning loop that provides causal context for bug-fixing by monitoring simulation paths and internal state transitions. Furthermore, we integrate a Concolic Testing engine for systematic corner-case exploration and provide a granular assessment of our dead-code pruning mechanism. Finally, the experimental section has been significantly extended with additional experiments and ablation studies.

The remainder of this paper is organized as follows. Section~\ref{subsec:background} reviews the related work and provides the necessary background. Section~\ref{subsec:method} introduces the architecture of the AutoVeriFix+ framework with the Trace-Aware Optimization.  
Section~\ref{subsec:experiment} presents experimental results, and Section~\ref{subsec:conclusion} concludes the paper.

\section{BACKGROUND AND PRIOR WORK}
\label{subsec:background}

\subsection{Advances in Large Language Models}
In recent years, large language models have achieved significant breakthroughs, demonstrating exceptional capabilities in language understanding and generation~\cite{sun2025surveyneuralcodeintelligence, Nijkamp_Hayashi_Xiong_Savarese_Zhou,Mastropaolo_Pascarella_Guglielmi_Ciniselli_Scalabrino_Oliveto_Bavota_2023}. Contemporary LLMs, such as ChatGPT~\cite{openai2023gpt35} and LLaMA~\cite{touvron2023llamaopenefficientfoundation}, are equipped with billions of parameters and are trained on extensive datasets from diverse domains.

One of the most widely-used applications of LLMs is automated code generation. Models like OpenAI's Codex~\cite{openai2021codex} and DeepMind's AlphaCode~\cite{Li_2022} have demonstrated their ability to convert natural language descriptions into executable code. These models are trained on extensive programming datasets and are capable of performing tasks such as code completion, debugging, and solving complex programming challenges. By automating such tasks, LLMs are revolutionizing software development, boosting productivity, and lowering the entry barrier for programming. However, while LLMs excel in software programming languages like Python, C++, and Java, extending their capabilities to hardware description languages like Verilog remains challenging~\cite{Chip-Chat,gai2025exploringcodelanguagemodels,sun2025surveyneuralcodeintelligence}. Unlike software, hardware design requires precise specifications of timing, concurrency, and signal behaviors. Furthermore, the limited availability of high-quality, publicly accessible RTL datasets restricts the ability of general-purpose LLMs to master hardware-specific tasks effectively~\cite{Chip-Chat,chen2021codex,BetterV}. To bridge this gap, recent research has branched into two main directions: improving data quality through data-centric approaches and ensuring correctness via feedback-driven mechanisms.

\begin{figure*}
    \centering
    \includegraphics[width=0.95\linewidth]{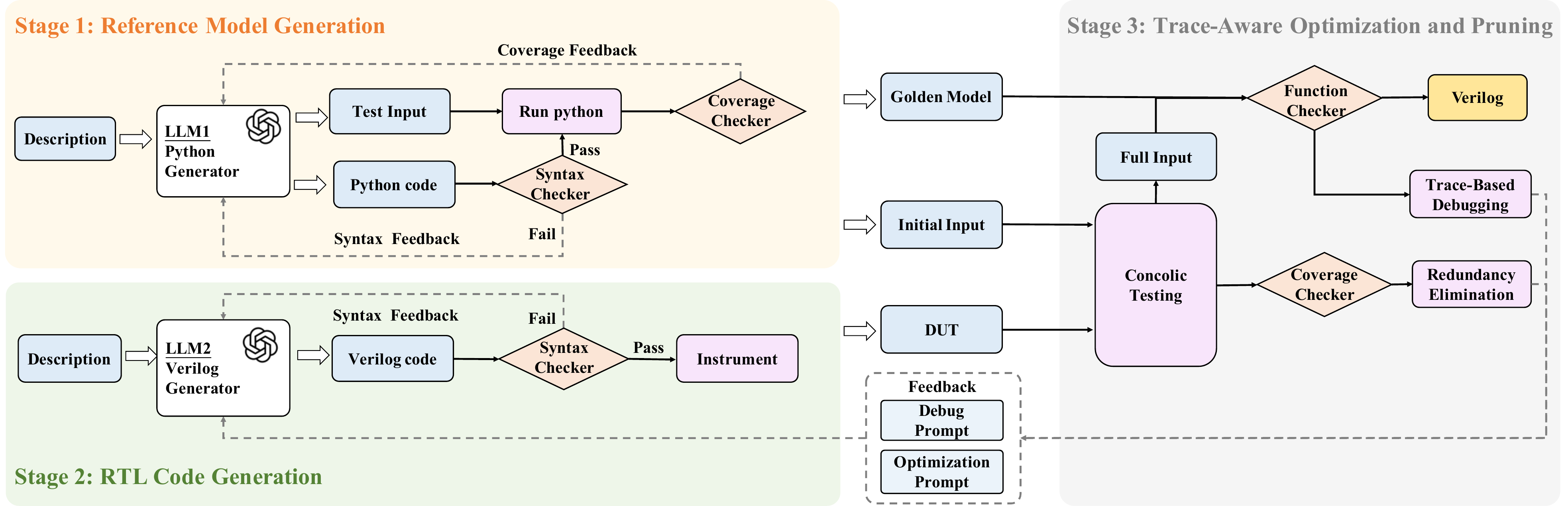}
    \caption{Overview of the three-stage AutoVeriFix+ framework. Stage 1 establishes a high-level Python golden model and initial test vectors. Stage 2 generates the initial Verilog RTL and iteratively repairs syntactic and functional errors. Stage 3 employs Concolic testing and trace-aware analysis to facilitate deep-path functional testing and redundancy elimination.}
    \vspace{-0.1in}
    \label{fig:overview}
\end{figure*}

\subsection{Data-Centric Approaches}
To address the data scarcity issue, researchers have focused on curating hardware-specific datasets and employing augmentation techniques to improve model performance. Thakur \textit{et al.}~\cite{thakur2024verigen} constructed an unsupervised Verilog dataset sourced from GitHub repositories and Verilog textbooks to fine-tune CodeGen-16B~\cite{nijkamp2022codegen}. However, due to the absence of manual annotations in this dataset, models fine-tuned on this dataset still underperform compared to commercial tools such as GPT-3.5. In contrast, VerilogEval~\cite{VerilogEval} developed a supervised dataset by carefully gathering, filtering, and refining Verilog files and design snippets from open-source hardware repositories. These rigorous steps ensure the inclusion of representative RTL code, enabling LLMs to learn hardware design patterns more effectively. Fine-tuned on the supervised dataset, VerilogEval became the first non-commercial model to match GPT-3.5 performance. 

Data augmentation techniques have been employed to further enhance dataset diversity. RTLCoder~\cite{Xie_2023} introduces a framework that generates natural language descriptions aligned with Verilog code semantics, creating contextual pairings that enrich the training process. Similarly, OriGen~\cite{2024origen} employs a code-to-code augmentation strategy, producing enhanced datasets that include both original designs and error-corrected versions of RTL code. HLSDebugger~\cite{wang2025hlsdebugger}, by training on a large-scale dataset containing 300,000 samples, utilizes an encoder-decoder structure to uniformly achieve localization, type prediction, and automatic repair of logical errors. 

Although these approaches enhance LLMs' comprehension of RTL design principles, they remain insufficient for ensuring syntactic and functional correctness. Specifically, they primarily focus on ``generation'' capabilities rather than satisfying the rigorous verification standards mandated by hardware design.

\subsection{Feedback-driven Mechanism}
To improve the correctness of generated code, feedback-driven mechanisms have been integrated into the code generation workflow in LLM-generated RTL code. The most common approach involves leveraging EDA tool logs to fix syntax errors. AutoChip~\cite{AutoChip} and OriGen~\cite{2024origen} pioneered the use of compiler error messages as prompts for self-correction, enabling LLMs to iteratively resolve syntax errors and linter warnings. RTLFixer~\cite{RTLFixer} further optimizes this process by using Retrieval-Augmented Generation (RAG) to retrieve debugging guidance related to specific compilation errors. 

Beyond simple compilation, some frameworks incorporate secondary metrics to guide the model. RTLCoder~\cite{Xie_2023} introduces a code quality scoring feedback mechanism, where the generated code is evaluated based on syntactic correctness and similarity to reference code. The scoring results are then used as feedback to enhance the model's generation capability. Similarly, some studies have integrated physical implementation feedback, specifically Power, Performance, and Area (PPA) metrics, to guide the LLM toward hardware-efficient implementations~\cite{BetterV,Chang_Wang_Ren_Wang_Liang_Han_Li_Li}.

Although these previous studies have laid the foundation for automatic code generation, we have observed significant functional errors in the Verilog code produced by these approaches, particularly in common hardware design tasks such as multi-branch logic and complex state transitions. Unlike earlier approaches that relied solely on compiler syntactic feedback, we use a reference model to generate high-quality testbenches and perform dynamic simulation to evaluate and improve the functional correctness of the code.

\section{METHODOLOGY}
\label{subsec:method}

\subsection{Overview}
\label{subsec:overview}
To comprehensively improve functional correctness and implementation efficiency, we propose AutoVeriFix+, a three-stage feedback framework designed to automatically detect, correct, and optimize LLM-generated Verilog code. The core philosophy of AutoVeriFix+ is to leverage the high functional accuracy of LLMs in high-level languages (Python) to oversee the more error-prone process of hardware description language (Verilog) generation. As demonstrated in Section~\ref{subsec:refeval}, while LLMs achieve over 95\% functional correctness in Python, their Verilog output often falls below 50\% due to the scarcity of high-quality RTL training data. Ideally, using the generated Python code as a near-perfect reference model can elevate the functional correctness of LLM-generated Verilog code to levels comparable with that of the reference model. As illustrated in Fig.~\ref{fig:overview}, AutoVeriFix+ integrates LLMs with a concolic testing engine in a closed-loop framework designed to automate the generation, verification, and optimization of Verilog code.


\subsection{Stage 1: Python-Assisted Reference Modeling}
\label{subsec:refmodel}
In the first stage, the Python Generator (LLM1) translates the natural-language hardware specification into a high-level Python reference model and a corresponding set of initial test vectors. This process establishes a functional golden model that serves as the ground truth for all subsequent verification cycles. Unlike conventional HLS flows that convert high-level Python or C++ into Verilog, our framework uses Python code solely as a functional oracle. 

To ensure the reliability of this oracle, the generated Python code is subjected to a two-fold validation process:
\begin{itemize}
    \item Syntax and interface validation: The generated code is processed by a syntax checker to ensure it is executable and conforms to the interface. If compilation errors or interface mismatches are detected, the framework initiates a syntax feedback loop to fix them.
    \item Coverage analysis: The reference model is executed to assess logic branch coverage. If coverage is insufficient, a feedback loop triggers LLM1 to expand the test inputs, ensuring a robust starting point for downstream concolic exploration.
\end{itemize}

\subsubsection{Python Generation and Syntax Validation} The process begins by providing a structured functional description of the hardware design to LLM1. The model is prompted to produce two artifacts: (1) a functional Python Reference Model embodying the design logic, and (2) a set of Initial Test Inputs for verification. 

The generated Python code is immediately processed by a Syntax Checker. In addition to verifying executability, this checker validates the consistency of the interface between the test inputs and the model's port definitions. If discrepancies or syntax errors are identified, the framework invokes an iterative \textit{Syntax Feedback Loop}. This loop returns specific error logs and diagnostic messages to LLM1, compelling the model to resolve structural or logical violations until a stable, executable reference is achieved.

\begin{figure}[t]
    \centering
    \includegraphics[width=0.98\linewidth]{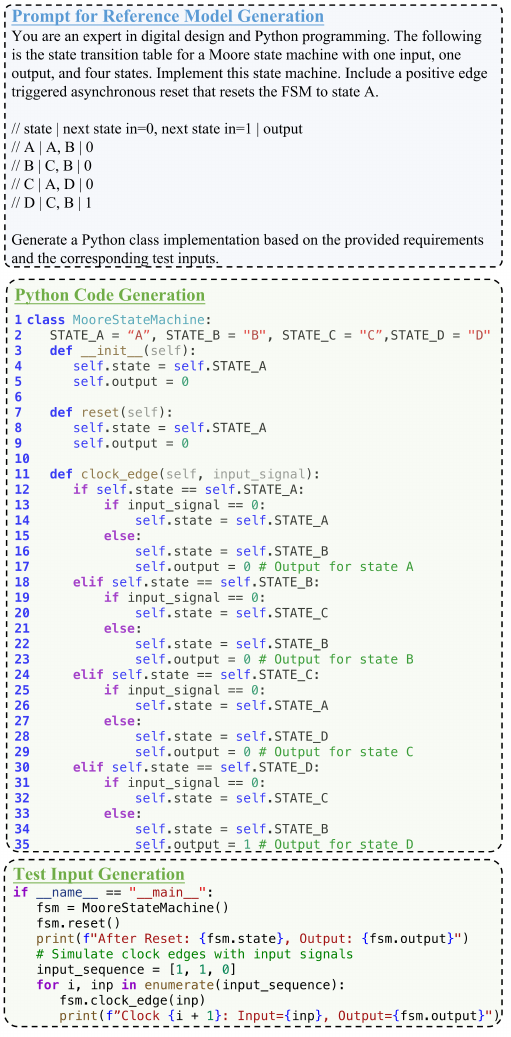}
    \caption{Our framework provides the LLM with a structured functional description of the hardware design, and the LLM generates the Python class and the initial test sequence.}
    \label{fig:reference}
\end{figure}

In the simple example illustrated in Fig.~\ref{fig:reference}, we start by giving LLM1 a structured description of a state machine (shown at the top of Fig.~\ref{fig:reference}). LLM1 then generates a Python class that captures the behavior of the state machine (illustrated in the middle of Fig.~\ref{fig:reference}) and an initial input sequence of [1,1,0] (shown at the bottom of Fig.~\ref{fig:reference}). Once the Python class and test inputs are created, they undergo a syntax check. If syntax errors are detected, the framework provides feedback in the form of error messages, allowing the LLM to adjust its output as necessary.

In the example provided in Fig.~\ref{fig:reference}, the generated reference model effectively captures the functional behavior of the hardware. This success can be attributed to the extensive training of LLMs on large Python datasets, which enables them to produce highly accurate and readable Python implementations that align with the provided functional descriptions.

\subsubsection{Coverage-Driven Test Input Refinement}

Once the Python code is syntactically valid, the framework executes the model using the initial inputs. The execution trace is analyzed by a Coverage Checker to assess how thoroughly the logic branches are exercised. If the branch coverage falls below a predefined threshold (e.g., 85\%), the checker generates a Coverage Message detailing uncovered regions. This feedback guides LLM1 to generate more targeted test vectors.

It is important to note that this threshold is loosely defined and serves primarily as an optimization for Stage 3. The objective is to provide high-quality starting seeds for the concolic testing engine. While the concolic engine is expected to achieve high coverage even from low-quality initial inputs, providing a high-coverage seed set in Stage 1 significantly reduces the computational time and solver iterations required in Stage 3 to explore deep logic states.

\begin{figure}[h]
    \centering
    \includegraphics[width=0.98\linewidth]{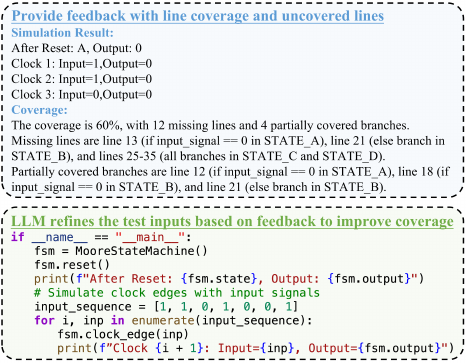}
    \caption{LLM refines the test inputs with coverage feedback}
    \label{fig:refine}
\end{figure}

For example, as illustrated in Fig.~\ref{fig:refine}, the initial coverage is 60\%, leaving 4 branches uncovered. The framework identifies these specific gaps, prompting the LLM to generate refined vectors (e.g., [1,1,0,1,0,0,1]). This iterative refinement continues until the system outputs a finalized Golden Model and a high-coverage input set, ensuring the downstream concolic-driven Stage 3 is efficient.

\subsection{Stage 2: Iterative RTL Refinement}
\label{subsec:rtlgen}
Stage 2 focuses on translating the functional specifications into a synthesizable Register Transfer Level (RTL) implementation. This stage operates in a closed loop that handles initial generation, syntax repair, and automated instrumentation, preparing the design for the deep functional analysis performed in Stage 3.

Existing LLM-based hardware generation processes often rely on black-box functional verification, which checks only I/O port correspondence. This approach presents significant limitations for production-quality hardware. First, silent bugs can occur when internal state transitions are incorrect, yet the primary outputs happen to match the expected values on a limited test set. Second, LLMs frequently hallucinate redundant logic or dead code that passes functional simulation but degrades the design's Area, Power, and Timing metrics. To address these challenges, Stage 2 applies subsequent refinement informed by feedback from Stage 3.

\subsubsection{Verilog Generation and Syntax Validation}

\begin{figure}
    \centering
    \includegraphics[width=0.98\linewidth]{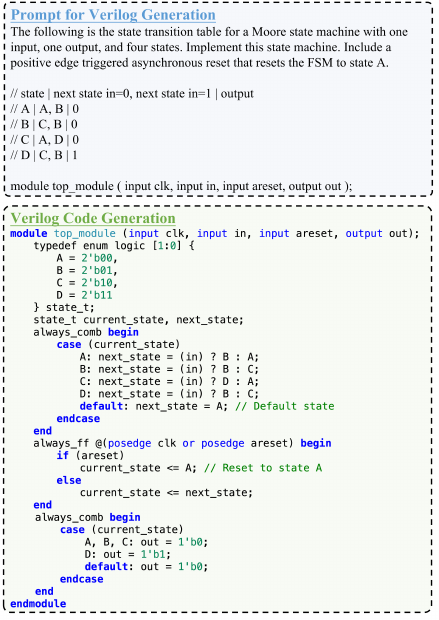}
    \caption{An illustrative example of prompt engineering for Verilog generation.}
    \label{fig:verilog}
\end{figure}

The Verilog Generator (LLM2) generates an initial Verilog code based on the structured hardware description. In addition to the detailed behavioral description provided in Stage 1, the prompt to LLM2 also includes comprehensive design specifications, such as module definitions and I/O port configurations. LLM2 then generates a complete Verilog implementation that reflects the specified design requirements, as shown in Fig.~\ref{fig:verilog}, illustrating the model's capability to translate natural language specifications into structured RTL designs.

\begin{figure}[t]
    \centering
    \includegraphics[width=0.95\linewidth]{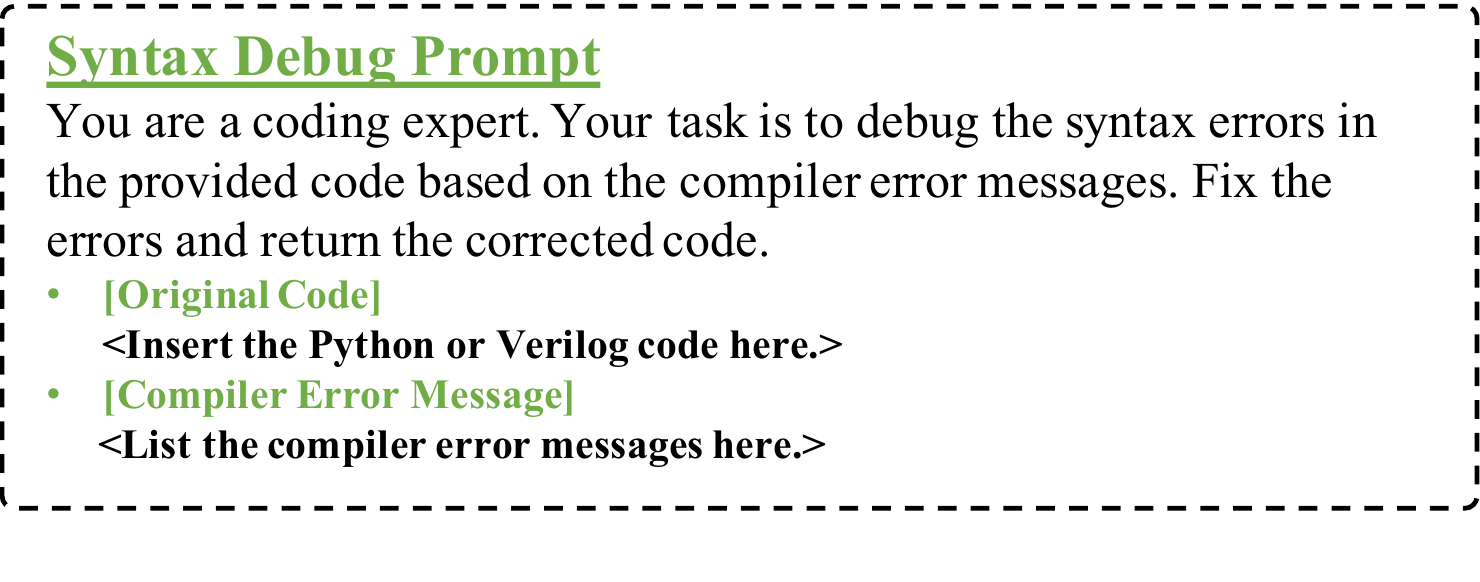}
    \caption{The standardized Syntax Debug Prompt template}
    \label{fig:debug}
\end{figure}

Similar to Stage 1, the generated Verilog undergoes a rigorous syntax check to ensure it compiles correctly. If the hardware compiler (e.g., Icarus Verilog or Verilator) detects structural violations, such as undeclared signals, missing semicolons, or keyword misplacements, the framework automatically generates a Syntax Debug Prompt (Fig.~\ref{fig:debug}). This prompt explicitly pairs the original buggy code with the compiler’s diagnostic messages, including error types, specific line locations, and additional diagnostic metadata. By iteratively feeding this context back into LLM2, the framework compels the model to analyze failures and correct syntax violations. This iterative repair cycle continues until the design achieves successful compilation, ensuring the Verilog code is syntactically sound and ready for the subsequent instrumentation.

\subsubsection{RTL Code Instrumentation}

To enable precise debugging and hardware execution logic tracking, we implement an automated instrumentation pass that transforms the raw Verilog into a Design Under Test (DUT), which is then forwarded to Stage 3. The goal of this process is to capture the concrete execution path and the instantaneous register states during simulation, enabling the detection of deep functional divergences that black-box testing cannot detect.

The instrumentation workflow proceeds as follows:
\begin{itemize}
    \item The framework parses the LLM-generated Verilog into an Abstract Syntax Tree (AST) to construct a Control Flow Graph (CFG).
    \item Each branch within the CFG is assigned a unique identifier ($Bi$). The framework then injects monitoring logic (e.g., \texttt{display} statements) at the end of each basic block.
    \item Finally, the framework identifies all the registers and inserts display statements to log the state of every register at each clock cycle.
\end{itemize}

\begin{figure}[htpb]
\begin{minipage}{\linewidth}
\begin{lstlisting}[caption={Example 1}, label={lst:example1}]
module top (
    input  wire       clock,
    input  wire       reset,
    input  wire [7:0] in,   
    output reg        out   
);

    reg [7:0] counter;  

    always @(posedge clock) begin
        if (reset) begin
            counter <= 8'h00; 
            out <= 1'b0; 
            $display("B_1");
        end
        else begin
            case (in)
                8'h00: begin
                    counter <= counter - 1; 
                    $display("B_2");
                end
                8'h02: begin
                    counter <= counter + 1;
                    $display("B_3");
                end
                8'hFF: begin
                    counter <= 8'h00; 
                    $display("B_4");
                end
                default: begin
                    $display("B_5");
                end
            endcase
            if (counter == 8'h01) begin
                out <= 1'b1; $display("B_6");
            end
            else begin
                out <= 1'b0; $display("B_7");
            end
        end
    end
    
    // Display the registers
    always @(posedge clock) begin
        $display("R counter = %d", counter);
        $display("R out = %d", out);
    end
endmodule

\end{lstlisting}
\end{minipage}
\end{figure}

The example of an instrumented design is shown in Listing 1 with the injected monitoring statements. During simulation, the injected code generates a runtime execution trace and periodic state snapshots.  By comparing this concrete path against the high-level logic flow of the Python reference model, the framework can pinpoint the exact cycle and location where the hardware logic deviates from the intended behavior.

\subsubsection{Optimization Loop}

However, achieving syntactic correctness does not guarantee functional correctness or implementation efficiency. Once instrumented, the DUT advances to Stage 3 for rigorous functional evaluation and optimization. As illustrated in the framework overview (Fig.~\ref{fig:overview}), the interaction between Stage 2 and Stage 3 forms a closed-loop refinement process. When the downstream analysis detects functional discrepancies or identifies unreachable logic, it generates high-level semantic feedback in the form of Trace-based Debug prompts (Fig.~\ref{fig: debug}) and Redundancy Elimination prompts (Fig.~\ref{fig: optimize}). These signals provide the LLM with detailed, causal context regarding internal state failures and implementation inefficiencies, as discussed in depth in Section~\ref{subsec:rtloptimaze}. By iteratively processing this feedback, LLM2 refines the Verilog implementation, incrementally fixing logic bugs and pruning dead code until the final RTL design is both functionally equivalent to the Golden Model and optimized for hardware implementation.




\subsection{Stage 3: Trace-Aware Optimization and Pruning}
\label{subsec:rtloptimaze}
While the Verilog code generated in Stage 2 is syntactically valid, it does not guarantee functional correctness or hardware efficiency. LLM-generated hardware descriptions often suffer from hallucinations, where logic deviates from specifications due to subtle timing issues or state-transition discrepancies. Furthermore, LLMs frequently produce redundant logic that passes basic simulations but results in excessive area overhead.

To address these challenges, we apply trace-aware optimization and pruning with a concolic-based verification mechanism. This stage moves beyond traditional black-box verification (which monitors only I/O ports) by systematically exploring the state space and aligning the internal execution trace of the Verilog code with the Python reference model. With the instrumented DUT from Stage 2, our framework detects the exact clock cycle and path trace at which the internal hardware state diverges from the golden model. This provides LLM2 with granular causal context for bug fixing rather than generic I/O mismatch messages. Furthermore, by identifying functionally unreachable code blocks, we can prune redundant logic to optimize the hardware area without compromising functionality. 

As shown in the right part of Fig.~\ref{fig:overview}, Stage 3 takes the Python Reference Model and the initial high-quality input test produced in Stage 1, and the DUT generated in Stage 2. Although the input vectors from Stage 1 provide a solid functional baseline, they often fail to exercise deep sequential logic or corner cases in the hardware implementation. Therefore, Stage 3 is built upon a Concolic testing engine to expand the initial input into a test set with much higher coverage. A function checker then compares the DUT outputs against those of the golden model under the expanded input set. When discrepancies are detected, a trace-based debugging module performs causal trace analysis and generates a corresponding debugging message. In parallel, a coverage checker evaluates the reachability of logic branches, flags unreachable code, and produces a coverage message. Then, both the debugging and coverage messages are fed back to the LLM in Stage 2, forming a closed-loop optimization process that incrementally guides the model to refine the Verilog code until it achieves functional correctness while eliminating redundant logic.

\subsubsection{Concolic Testing}

The core engine of Stage 3 is a Concolic testing framework (combining \textit{concrete} simulation and \textit{symbolic} execution) adapted from~\cite{concolic}. This engine targets the verification bottleneck by systematically mutating branches in simulation paths to exercise deep logic that extensive random testing might overlook. While Concolic testing cannot guarantee 100\% functional correctness, experimental results demonstrate that our framework effectively corrects a large portion of errors in LLM-generated Verilog code.

\begin{figure}[t]
    \centering
    \includegraphics[width=0.75\columnwidth]{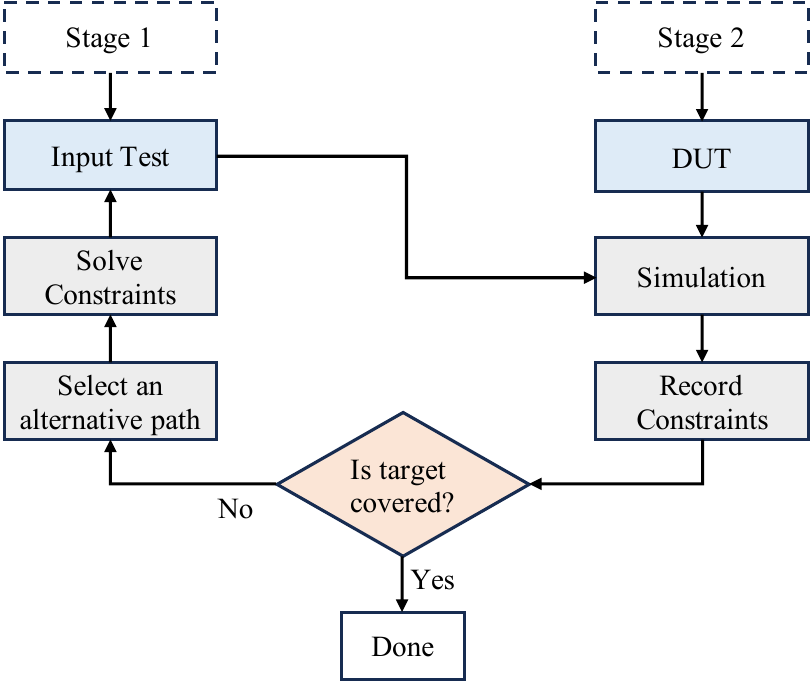}
    \caption{The Workflow of a Concolic Testing Engine}
    \label{fig: Traditional Concolic Framework}
\end{figure}

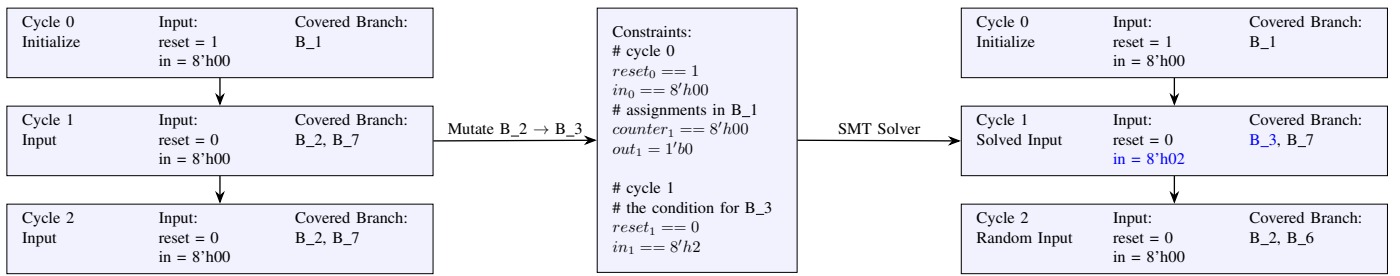
\begin{figure*}[b]
    \centering
    \resizebox{\linewidth}{!}{ 
        \begin{tikzpicture}[
    node distance=0.6cm,
    block/.style={
        draw,
        fill=blue!5, 
        rectangle,
        minimum width=4cm,
        minimum height=1.5cm,
        align=left
    },
    arrow/.style={
        thick,
        -{Stealth[scale=1.2]}
    }
]

    
    \node[block] (c0) {
        \begin{tabular}{p{2.5cm} p{2.5cm} p{2.5cm}}
        Cycle 0 & Input: & Covered Branch: \\
        Initialize & reset = 1 & B\_1 \\
        & in = 8'h00 & 
        \end{tabular}
    };

    \node[block, below=of c0] (c1) {
        \begin{tabular}{p{2.5cm} p{2.5cm} p{2.5cm}}
        Cycle 1 & Input: & Covered Branch: \\
        Input & reset = 0 & B\_2, B\_7 \\
        & in = 8'h00 & 
        \end{tabular}
    };

    \node[block, below=of c1] (c2) {
        \begin{tabular}{p{2.5cm} p{2.5cm} p{2.5cm}}
        Cycle 2 & Input: & Covered Branch: \\
        Input & reset = 0 & B\_2, B\_7 \\
        & in = 8'h00 & 
        \end{tabular}
    };


    \node[block, minimum height=5.7cm, right=3.5cm of c0.north east, anchor=north west] (cc0) {
        \begin{tabular}{p{3.5cm}}
        Constraints: \\
        \# cycle 0\\
        $reset_0 == 1$\\
        $in_0 == 8'h00$\\
        \# assignments in B\_1\\
        $counter_1 == 8'h00$\\
        $out_1 = 1'b0$ \\
            \\
        \# cycle 1\\
        \# the condition for B\_3\\
        $reset_1 == 0$\\
        $in_1 == 8'h2$
        \end{tabular}
    };

    \node[block, right=3.5cm of cc0.north east, anchor=north west] (ccc0) {
        \begin{tabular}{p{2.5cm} p{2.5cm} p{2.5cm}}
        Cycle 0 & Input: & Covered Branch: \\
        Initialize & reset = 1 & B\_1 \\
        & in = 8'h00 & 
        \end{tabular}
    };

    \node[block, below=of ccc0] (ccc1) {
        \begin{tabular}{p{2.5cm} p{2.5cm} p{2.5cm}}
        Cycle 1 & Input: & Covered Branch: \\
        Solved Input & reset = 0 & \textcolor{blue}{B\_3}, B\_7 \\
        & \textcolor{blue}{in = 8'h02} & 
        \end{tabular}
    };

    \node[block, below=of ccc1] (ccc2) {
        \begin{tabular}{p{2.5cm} p{2.5cm} p{2.5cm}}
        Cycle 2 & Input: & Covered Branch: \\
        Random Input & reset = 0 & B\_2, B\_6 \\
        & in = 8'h00 & 
        \end{tabular}
    };


    \draw[arrow] (c0.south) -- (c1.north);
    \draw[arrow] (c1.south) -- (c2.north);

    \draw[arrow] (c1.east) -- (cc0.west) node[midway, above] {Mutate B\_2 $\rightarrow$ B\_3};
    \draw[arrow] (cc0.east) -- (ccc1.west) node[midway, above] {SMT Solver};

    \draw[arrow] (ccc0.south) -- (ccc1.north);
    \draw[arrow] (ccc1.south) -- (ccc2.north);
\end{tikzpicture}
    }
    \caption{An illustrative example of branch mutation in Concolic testing. The Concolic engine generates mutated inputs to guide simulation toward uncovered branches.}
    \label{fig: Concolic Mutate Example}
\end{figure*}

The workflow, depicted in Fig.~\ref{fig: Traditional Concolic Framework}, follows a path-exploration strategy to expand the simulation boundary. The process begins by simulating the DUT with the high-quality input test from Stage 1, enabling the Concolic engine to achieve coverage convergence more quickly than with random inputs. Utilizing the monitoring logic injected during Stage 2, the engine records real-time simulation paths and cycle-accurate state snapshots. If coverage is improved during the simulation, the input vector is added to the input set. Otherwise, the engine heuristically selects a branch node along the current path for mutation. It then constructs a set of symbolic constraints representing the negated path condition required to force the execution flow into the selected branch. These constraints are fed to an SMT solver. If they are satisfiable, the solver returns a new input vector that is guaranteed to drive the subsequent simulation iteration toward the selected branch. By iteratively updating constraints and generating directed inputs, the framework progressively penetrates deep logic states. This process ensures that the testbench evolves from merely checking high-level logic to covering as many branches as possible in the Verilog implementation.

\subsubsection{Branch Mutation in Concolic Testing}
To visualize the underlying mechanics of state-space traversal, Fig.~\ref{fig: Concolic Mutate Example} illustrates a single iteration of the branch mutation process based on the RTL implementation provided in Listing~\ref{lst:example1}.

\textbf{Simulation and Constraint Extraction:}
The process begins with an initial input sequence (e.g., $\{in=8'h00\}$ for all three cycles) generated in Stage 1, starting with initialization $\{reset=1\}$. As shown in the trace, this default sequence limits execution to a repetitive path encompassing branches [B\_1, B\_2, B\_7, B\_2, B\_7].

\textbf{Branch Mutation and Constraint Solving:}
To increase coverage and expand the input set, the Concolic engine identifies an unexplored branch, e.g., by mutating branch B\_2 to branch B\_3. As shown in Fig.~\ref{fig: Concolic Mutate Example}, the engine extracts the cumulative path constraints from Cycle 0 up to the target junction and appends the logical predicate required to satisfy B\_3. By solving the constraints using an SMT solver, the system generates a new, directed input vector $\{reset=0, in=8'h02\}$. In the subsequent simulation, this mutated vector successfully triggers branch B\_3. Furthermore, as the counter increments within B\_3, the precondition for branch B\_6 is subsequently satisfied, leading to an immediate gain in functional coverage.

This single mutation cycle demonstrates the framework's capability to resolve multi-cycle sequences necessary to reach deep logic corners. AutoVeriFix+ leverages this capability to generate a comprehensive test suite, designated the Full Input set, that maximizes functional observability of the generated RTL.

\subsubsection{Trace-Based Functional Debugging}
The Function Checker executes the DUT and the Python Golden Model in parallel using the Full Input set. In a traditional differential testing paradigm, the system only detects an external output mismatch ($Output_{DUT} \neq Output_{Golden}$). However, such black-box feedback is often insufficient for diagnosing complex hardware logic errors. 

\begin{figure}[t]
    \centering
    \includegraphics[width=0.95\columnwidth]{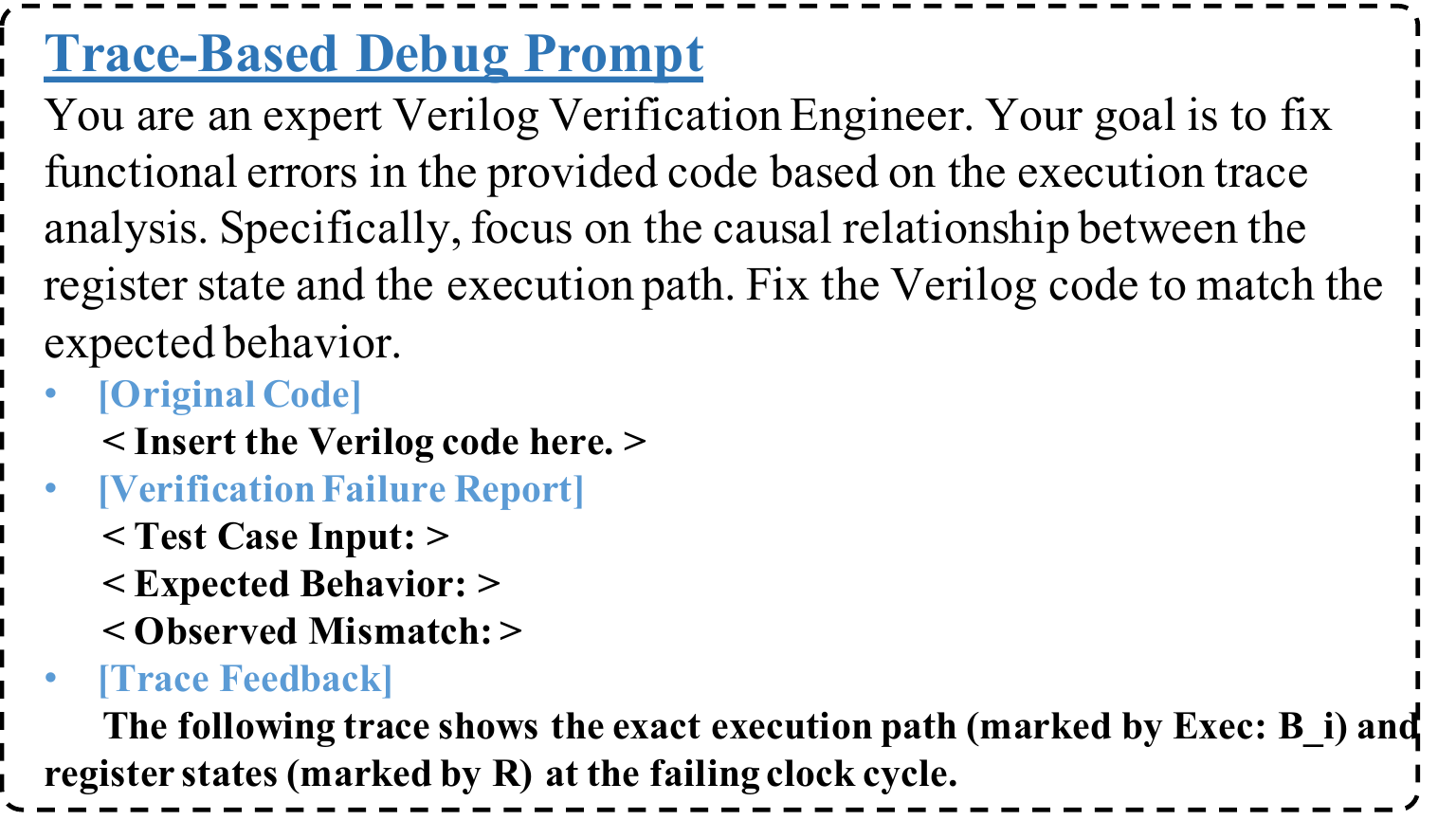}
    \caption{The Trace-Based Debug prompt for fixing functional errors.}
    \label{fig: debug}
\end{figure}

To address this observability gap, our framework employs a trace-based feedback mechanism. We utilize execution logs generated by tracking simulation paths and monitoring all internal register states to provide the LLM with deep causal context. Specifically, we construct a ``Trace-Based Debug Prompt'' which is composed of three primary components, as illustrated in Fig.~\ref{fig: debug}.

\begin{itemize}
    \item Original Code: The current version of the Verilog implementation under test.
    \item Verification Failure Report: A comprehensive report including the specific test case that triggered the failure, the expected output from the Golden Model, and the erroneous output observed from the DUT.
    \item Trace Feedback: The cycle-by-cycle execution trace, which consists of two parts for each cycle. 
    \begin{itemize}
    \item Execution Paths: The specific assignment identifiers (B\_{i}) triggered in each cycle.
    \item Internal State Snapshots: The instantaneous register states (R) at every clock edge.
\end{itemize}
\end{itemize}

This structured feedback compels the LLM to perform causal trace analysis, examining the relationship between internal register values and the resulting erroneous execution path. By enabling the model to trace signal propagation and state transitions cycle-by-cycle, the framework allows the LLM to pinpoint the exact temporal and structural point where the hardware state diverged from the functional expectation. 

Once the LLM generates a refined Verilog candidate, the updated code is re-integrated into the Concolic engine to regenerate the Full Input set. This step is crucial: a fix for one logic branch may break another or create new unreachable states. The design is then re-simulated, and outputs are reassessed against the golden model. This iterative loop of trace-based functional debugging continues until the observed hardware behavior matches exactly with the reference model under the Full Input test suite.

\subsubsection{Redundancy Elimination and Optimization}
\begin{figure}[t]
    \centering
    \includegraphics[width=0.95\columnwidth]{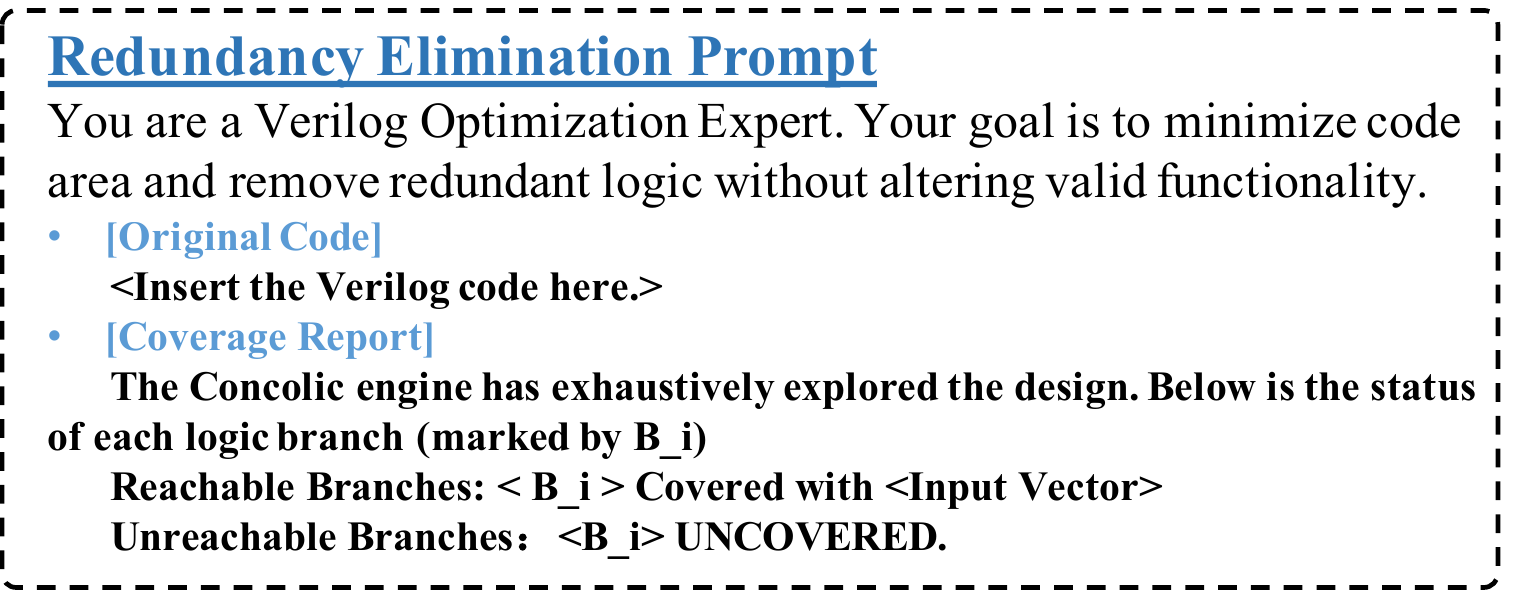}  
    \caption{The Redundancy Elimination prompt for removing dead code based on concolic coverage reports}
    \label{fig: optimize}
\end{figure}

In addition to causal debugging, Stage 3 also tries to refine the generated RTL design through a coverage-driven optimization. After Concolic testing completes, it classifies logic branches into two types based on the solver's results: reachable and potentially unreachable branches. To optimize the hardware area, we design the ``Redundancy Elimination'' prompt, as shown in Fig.~\ref{fig: optimize}. The LLM serves as an optimization expert for semantic logic pruning. It uses the reachability report to eliminate dead code without altering the functionality of reachable branches, ensuring that the final design is both correct and efficient.

While Concolic testing is significantly more effective at path exploration than random fuzzing, the inherent complexity of hardware state spaces means it may occasionally fail to reach certain valid logic branches due to solver timeouts or state-space explosion. Consequently, branches identified as unreachable are treated as candidates for speculative pruning. Notably, these candidates often include necessary defensive logic (e.g., FSM default states) that are inherently unreachable under normal operation but critical for hardware safety. Instead of computationally expensive formal verification ,  we leverage the LLM’s semantic understanding of the design’s intent to eliminate these potential redundancies. If a block is identified as redundant, the LLM performs a trial deletion, and the framework executes the debugging against the golden model in the next loop. This ensures that the behavior remains invariant and that no false dead logic is erroneously removed. In practice, high-level specifications are important for facilitating LLMs to generate initial inputs in Stage 1 and identify false dead logic during Stage 3 optimization.

\subsection{Summary}

In summary, our proposed framework establishes a novel paradigm for automated hardware generation by leveraging software-centric high-level reasoning to guide hardware-specific verification. By incorporating a software-to-hardware golden model reference and a white-box, trace-aware feedback loop, we effectively address the primary shortcomings of existing LLM-based RTL generation, which lacks adaptive functional feedback. Consequently, our approach significantly mitigates the risk of hardware bugs in LLM-generated designs while simultaneously optimizing them by pruning redundancy, thereby providing an automated approach toward reliable hardware design.

\section{Experiment}
\label{subsec:experiment}

\subsection{Experimental Setup}
We implemented AutoVeriFix+ using OpenAI's GPT-3.5~\cite{openai2023gpt35} and GPT-4~\cite{openai2023gpt4} as the LLM tools for code generation. These models demonstrate exceptional performance in handling Python-based problems, which makes them highly effective in creating reference models. 

To evaluate the performance of Verilog code generation, we selected two representative benchmarks: VerilogEval~\cite{VerilogEval} and RTLLM~\cite{lu2024rtllm}. These datasets were specifically designed to evaluate the functional correctness and quality of Verilog code generated by language models. The VerilogEval dataset was derived from Verilog programming tasks sourced from the HDLBits instructional website. These tasks were divided into two subsets: VerilogEval-Human, with 156 carefully hand-crafted problems, and VerilogEval-Machine, comprising 143 problems generated by GPT. The RTLLM dataset v1.1 contains a total of 29 designs, while RTLLM v2.0 offers 50 diverse Verilog tasks that closely resemble real-world RTL design scenarios, providing varying levels of difficulty to evaluate the performance of AutoVeriFix+ and other existing approaches.

\subsection{Correctness of the Python Reference Model}
\label{subsec:refeval}

To validate the effectiveness of the reference model generated in Stage 1, we first analyzed the final Python code using two metrics: syntactic correctness and functional correctness. Syntactic correctness assesses whether the generated code compiles successfully without syntax errors. Functional correctness evaluates whether the code produces the expected outputs upon execution.

Table~\ref{tab:reference_model_results} shows the syntactic and functional correctness of the Python reference model. Since the results for GPT-3.5 are similar, we have omitted them from the table. The experimental results demonstrate the effectiveness of generating reference models for various benchmarks, achieving near-perfect syntactic correctness and over 94\% functional correctness across all benchmarks. Especially for the VerilogEval-machine benchmark, the reference model achieved an impressive functional correctness of 98.60\%. These results demonstrate that LLMs are capable of translating high-level hardware specifications into an executable Python oracle. This oracle provides a robust foundation for subsequent differential testing and Concolic exploration.

While large language models can generate high-quality Python code, the initial test inputs generated without feedback may exhibit insufficient coverage, leaving corner behaviors under-specified. To mitigate this issue, we incorporate a coverage-driven feedback mechanism that feeds the execution coverage of the Python model back to the LLM to refine test inputs. By enforcing this Python coverage feedback loop, the quality of the generated test vectors improves dramatically. As shown in Fig.~\ref{fig:FPR}, relying purely on the baseline GPT-4 inputs results in a high False Positive Rate (FPR) of 28.40\% on the RTLLM v2.0 benchmark. However, after applying the Python coverage feedback, the FPR drops significantly to 7.60\%. This demonstrates that Stage 1 provides a highly robust and trustworthy foundation for the subsequent differential testing.

\subsection{Functional Correctness of Verilog Code}

Following VerilogEval~\cite{VerilogEval}, we use the metric \textit{pass@k} to evaluate the functional correctness of the generated Verilog code. \textit{pass@k} represents the expected probability of at least one answer passing the functional evaluation when randomly choosing $k$ answers from $n$ candidates:

\begin{equation}
    \text{pass@k} := \mathbb{E}_{\text{Problems}} \left[ 1 - \frac{\binom{n-c}{k}}{\binom{n}{k}} \right]
\end{equation}
where $n$ represents the total number of generated samples for a given problem, $c$ denotes the number of correct samples, and $k$ specifies the number of selected code samples. We set $n=10$ according to the original settings.

\begin{table}[t]
\renewcommand{\arraystretch}{1.5}
\centering
\caption{Evaluation of the reference model generated in Stage 1 using GPT-4.}
\setlength{\tabcolsep}{4mm}{
\begin{tabular}{l|c|c}
\toprule
\multirow{2}{*}{\textbf{Verilog Dataset}} 
  & \textbf{Syntactic} & \textbf{Functional} \\
    & \textbf{Correctness (\%)}& \textbf{Correctness (\%)}  \\ \midrule
\textbf{VerilogEval-human}    & 99.35   & 96.15    \\ 
\textbf{VerilogEval-machine}   & 100.00   & 98.60 \\
\textbf{RTLLM v1.1}             & 100.00    & 96.55     \\ 
\textbf{RTLLM v2.0}              & 98.00    & 94.00   \\ 
\bottomrule
\end{tabular}
\label{tab:reference_model_results}
} 
\end{table}

\begin{table*}[htbp]
\caption{Comparison of functional correctness on the VerilogEval benchmarks~\cite{VerilogEval} and the RTLLM benchmarks~\cite{lu2024rtllm}. The top scores ranked 1\textsuperscript{st}, 2\textsuperscript{nd}, and 3\textsuperscript{rd} in each column are highlighted in \colorbox[HTML]{FDE9E9}{Red}, \colorbox[HTML]{E0F7FA}{Blue}, and \colorbox[HTML]{E8F5E9}{Green}, respectively.}
\label{tab:comparison}
\centering
\renewcommand{\arraystretch}{1.5}
\setlength{\tabcolsep}{4pt}
\begin{tabular}{p{2.5cm}|p{4.5cm}|ccc|ccc|c|c}
\bottomrule
\rowcolor[HTML]{FFFFFF} 

{\multirow{3}{*}{\textbf{Category}}} & \multirow{3}{*}{\textbf{Model}} & \multicolumn{3}{c|}{\textbf{VerilogEval-human}} & \multicolumn{3}{c|}{\textbf{VerilogEval-machine}} & \textbf{RTLLM} & \textbf{RTLLM} \\

 &  &  \multicolumn{3}{c|}{\textbf{(\%)}} & \multicolumn{3}{c|}{\textbf{(\%)}} & \textbf{~v1.1 (\%)~} & \textbf{~v2.0 (\%)~} \\ \cline{3-10}

&  & \textbf{{pass@1}} & \textbf{pass@5} & \textbf{pass@10} & \textbf{{pass@1}} & \textbf{pass@5} & \textbf{pass@10} & \textbf{pass@5} & \textbf{pass@5} \\  \midrule

\multirow{5}{2.5cm}{\textbf{Verilog-Specific Models}} 
 & VerilogEval~\cite{VerilogEval} & 28.8 & 45.9 & 52.3 & 46.2 & 67.3 & 73.7 & - & -\\  
   & CodeGen-6B MEV-LLM~\cite{nadimi2024multi} & 42.9 & 48.0 & 54.4 & 57.3 & 61.5 & 66.4 & - & -\\ 
& BetterV-CodeQwen~\cite{BetterV} & 46.1 & 53.7 & 58.2 & 68.1 & 79.4 & 84.5 & - & -\\ 
  & RTLCoder~\cite{Xie_2023} & 41.6 & 50.1 & 53.4 & 61.2 & 76.5 & 81.8 & 48.3 & - \\ 
 & OriGen~\cite{2024origen} & \cellcolor[HTML]{E8F5E9}54.4 & 60.1 & 64.2 & \cellcolor[HTML]{E0F7FA}74.1 & \cellcolor[HTML]{E0F7FA}82.4 & \cellcolor[HTML]{E0F7FA}85.7 & \cellcolor[HTML]{E8F5E9}65.5 & - \\ \hline

\multirow{3}{*}{\textbf{Open Source Models}} 
 & CodeLlama-7B-Instruct~\cite{roziere2023code} & 18.2 & 22.7 & 24.3 & 43.1 & 47.1 & 47.7 & 34.5 & 33.1\\ 
& CodeQwen1.5-7B-Chat~\cite{bai2023qwen} & 22.4 & 41.1 & 46.2 & 45.1 & 70.2 & 77.6 & 37.9 & 36.4 \\ 
& DeepSeek-Coder-7B-Instruct-v1.5~\cite{guo2024deepseek} & 31.7 & 42.8 & 46.8 & 55.7 & 73.9 & 77.6 & 37.9 & 36.4\\ \hline

\multirow{5}{*}{\textbf{Commercial LLM}} 
& Claude3-Sonnet~\cite{claude3_family} & 46.1 & 56 & 60.3 & 58.4 & 71.8 & 74.8 & 58.6 & 54.4 \\
& GPT-3.5~\cite{openai2023gpt35} & 35.6 & 48.8 & 52.6 & 49.4 & 72.7 & 77.6 & 44.8 & 36.2\\ 
& GPT-4~\cite{openai2023gpt4} & 43.5 & 55.8 & 58.9 & 60 & 70.6 & 73.5 & \cellcolor[HTML]{E8F5E9}65.5  & 58.7\\ 
& GPT-4 Turbo~\cite{openai2023gpt4} & 54.2 & \cellcolor[HTML]{E8F5E9}68.5 & \cellcolor[HTML]{E8F5E9}72.4 & 58.6 & 71.9 & 76.2 & \cellcolor[HTML]{E8F5E9}65.5 & \cellcolor[HTML]{E8F5E9}63.4\\  \hline
 
\multirow{2}{*}{\textbf{Ours}} 
 & AutoVeriFix+ with GPT-3.5 & \cellcolor[HTML]{E0F7FA}58.5 & \cellcolor[HTML]{E0F7FA}71.8 & \cellcolor[HTML]{E0F7FA}73.7 & \cellcolor[HTML]{E8F5E9}69.5 & \cellcolor[HTML]{E8F5E9}77.6 & \cellcolor[HTML]{E8F5E9}79.7 & \cellcolor[HTML]{E0F7FA}75.9 & \cellcolor[HTML]{E0F7FA}71.9\\
& AutoVeriFix+ with GPT4 & \cellcolor[HTML]{FDE9E9}78.1 & \cellcolor[HTML]{FDE9E9}84.4 & \cellcolor[HTML]{FDE9E9}87.2 & \cellcolor[HTML]{FDE9E9}83.7 & \cellcolor[HTML]{FDE9E9}89.2 & \cellcolor[HTML]{FDE9E9}91.6 & \cellcolor[HTML]{FDE9E9}90.2 & \cellcolor[HTML]{FDE9E9}85.4 \\ \bottomrule
\end{tabular}
\end{table*}

We compared our approach with multiple categories of models: 
\begin{itemize}
    \item Closed-source commercial LLMs: GPT-3.5~\cite{openai2023gpt35}, GPT-4~\cite{openai2023gpt4} and Claude3-Sonnet~\cite{claude3_family}.
    \item General open-source models: CodeLlama~\cite{roziere2023code}, CodeQwen~\cite{bai2023qwen}, and DeepSeek-Coder~\cite{guo2024deepseek}.
    \item State-of-the-art Verilog-specific models: VerilogEval~\cite{VerilogEval}, CodeGen-6B MEV-LLM~\cite{nadimi2024multi}, BetterV-CodeQwen~\cite{BetterV}, RTLCoder~\cite{Xie_2023} and OriGen~\cite{2024origen}.
\end{itemize}

Table~\ref{tab:comparison} presents the functional correctness results across the benchmarks. The experimental results indicate that AutoVeriFix+ achieves significant improvements not only over leading commercial LLMs, such as GPT-4 and Claude3, but also domain-specific models, including OriGen and RTLCoder. Even when utilizing GPT-3.5 within our framework, we outperform most models. When employing GPT-4 within our framework, our approach achieves greater performance, consistently outperforming all other models across all benchmarks.

\subsubsection{Performance on RTLLM v2.0} In the RTLLM v2.0 benchmark, AutoVeriFix+ achieves remarkable performance, scoring 71.9 with GPT-3.5 and 85.4 with GPT-4, significantly outperforming all baseline models. For instance, compared to the original GPT-3.5, AutoVeriFix+ with GPT-3.5 improves the score on RTLLM v2.0 (pass@5) by 98.6\% (from 36.2 to 71.9). This underscores the efficacy of the stage-based refinement process in resolving functional errors.

\subsubsection{Evaluation on VerilogEval} On the VerilogEval-human benchmark, AutoVeriFix+ with GPT-4 achieves a pass@10 score of 87.2, outperforming both commercial model GPT-4 Turbo (72.4) and Claude3-Sonnet model (60.3). The highest score recorded by AutoVeriFix+ was on the VerilogEval-machine benchmark, with a pass@10 score of 91.6, exceeding domain-specific models such as OriGen (85.7) and RTLCoder (81.8). This score is also quite close to the ideal performance of the reference model, as shown in Table~\ref{tab:reference_model_results}.

Overall, these results indicate that AutoVeriFix+ with GPT-4 not only outperforms commercial general-purpose LLMs but also surpasses domain-specific models in generating Verilog code with functional correctness.

\subsection{Ablation Experiment: Importance of Feedback Mechanisms}


To assess the individual contributions of each feedback mechanism within AutoVeriFix+, we conducted a comprehensive ablation experiment across the RTLLM v2.0 and VerilogEval benchmarks. The results, summarized in Table~\ref{tab:syntax1} and ~\ref{tab:syntax2}, illustrate how Syntax Feedback, I/O Functional Feedback, and Trace-Based Debugging progressively enhance code generation quality.

\subsubsection{Evaluation of Syntactic Feedback}
We first assessed the impact of the Stage 2 Syntax Feedback loop by comparing the zero-shot baseline of GPT-4 against a version augmented solely with syntactic feedback. As shown in Table~\ref{tab:syntax1}, the inclusion of automated compiler feedback significantly enhances the syntactic validity of the generated RTL implementations, increasing the pass@1 score from 58.6\% to 77.8\%. 

However, the impact of syntax repair on functional correctness varies by dataset complexity. On the simpler VerilogEval benchmarks (Table~\ref{tab:syntax2}), fixing syntax errors leads to a substantial jump in functional pass rates (e.g., the pass@1 score was increased from 43.5\% to 68.1\% on VerilogEval-Human). This suggests that for simpler problems, the initial LLM logic is often correct, and syntax errors are the primary barrier. Conversely, on the complex RTLLM v2.0 dataset, syntax repair alone yields negligible functional improvement (51.6\% to 52.4\%). This indicates that for complex hardware logic, compilation is a necessary but insufficient condition for correctness; the underlying logic remains flawed and requires semantic guidance.

\begin{table}[t]
\caption{The effect of syntactic feedback and functional feedback in improving the pass@k score on RTLLM v2.0.}
\label{tab:syntax1}
\centering
\renewcommand{\arraystretch}{1.1}
\setlength{\tabcolsep}{1.3pt} 
\begin{tabular}{p{2.7cm}|ccc|ccc}
\toprule
\textbf{Model} & \multicolumn{3}{c|}{\textbf{Syntax (\%)}} & \multicolumn{3}{c}{\textbf{Function (\%)}} \\
\cmidrule(lr){2-4} \cmidrule(lr){5-7}
& {pass@1} & {pass@5} & {pass@10} &{pass@1} & pass@5 & pass@10 \\
\midrule
GPT-4 &58.6 &80.3 &84.8 &51.6 &58.7 &68.9 \\ \hline
GPT-4 + Syntax Feedback     & 77.8 & 89.6 & 90.0 & 52.4 & 64.5 & 70.7 \\
\hline
GPT-4 + I/O Feedback   & 89.8 & 92.0 & 94.0 & 74.6 & 83.5 & 88.0 \\ \hline
GPT-4 + Trace-Aware Debug & 92.9 & 94.0 & 96.0 & 79.0 & 85.4 & 88.6 \\
\bottomrule
\end{tabular}
\end{table}

\begin{table}[t]
\caption{The effect of syntactic feedback and functional feedback in improving the pass@k score on VerilogEval-Human.}
\label{tab:syntax2}
\centering
\renewcommand{\arraystretch}{1.1}
\setlength{\tabcolsep}{1.3pt} 
\begin{tabular}{p{2.7cm}|ccc|ccc}
\toprule
\textbf{Model} & \multicolumn{3}{c|}{\textbf{Syntax (\%)}} & \multicolumn{3}{c}{\textbf{Function (\%)}} \\
\cmidrule(lr){2-4} \cmidrule(lr){5-7}
& {pass@1} & {pass@5} & {pass@10} &{pass@1} & pass@5 & pass@10 \\
\midrule
GPT-4 & 78.5 & 87.2 & 89.1 & 43.5 & 55.8 & 58.9 \\ \hline
GPT-4 + Syntax Feedback     & 91.5 & 96.7 & 97.2 & 68.1 & 70.2 & 75.8 \\ \hline
GPT-4 + I/O Feedback   & 98.0 & 99.6 & 99.9 & 77.2 & 82.7 & 84.6 \\ \hline
GPT-4 + Trace-Aware Debug & 98.6 & 99.6 & 99.9 & 78.1 & 84.4 & 87.2 \\
\bottomrule
\end{tabular}
\end{table}

\subsubsection{Evaluation of I/O-Based Functional Feedback}
The integration of differential testing and the high-coverage Full Input suite in Stage 3 provides the most substantial improvement to functional correctness. To isolate the specific impact of the differential stimulus from the causal trace data, we conducted a controlled experiment. We constructed a restricted version of the ``Trace-Based Debug'' prompt (Fig.~\ref{fig: debug}) by omitting the Trace Feedback part (internal state snapshots and execution paths). This restricted prompt represents a black-box differential testing paradigm, which provides the LLM with only the specific failing test vector, the expected golden behavior, and the observed output mismatch.

The experimental results demonstrate that even without internal visibility, the high-quality test inputs generated by the Concolic engine significantly enhance the LLM's debugging capability. On RTLLM v2.0, the inclusion of I/O-based feedback drives the pass@1 functional score from 52.4\% to 74.6\%. This confirms that high-coverage differential testing against a golden reference is highly effective at resolving standard combinational and sequential logic errors. Similar improvements were observed on VerilogEval-Human (increasing from 68.1\% to 77.2\%), further validating the robustness and utility of the Python-based reference models in providing an accurate verification oracle for diverse hardware designs.

\subsubsection{Evaluation of Trace-Based Debugging Feedback}
Trace-Aware Debugging addresses complex functional bugs that cannot be effectively diagnosed using black-box I/O feedback alone. The effectiveness of trace-aware debugging is most evident on the challenging RTLLM v2.0 benchmark. While high-coverage I/O feedback establishes a strong baseline of 74.6\% pass@1 accuracy, the addition of fine-grained execution traces elevates performance to 79.0\%, representing a 4.4\% absolute gain. This result indicates that, for state-intensive designs, generic ``output mismatch'' information is insufficient for diagnosis. In contrast, cycle-accurate visibility into internal states and execution paths enables the LLM to perform causal reasoning and correct deep sequential and timing-related logic errors.


\subsection{Verification Integrity and False Positive Analysis}
\label{subsec:testeffect}
A significant challenge in automated RTL generation is the occurrence of false positives, in which an implementation is labelled ``correct’’ but contains internal functional defects. Unlike traditional generative approaches that provide no confidence measure, AutoVeriFix+ implements a transparent verification loop. By outputting differential simulation results alongside the final RTL, the framework provides a verification certificate, highlighting residual discrepancies for human-in-the-loop review.

\subsubsection{Classification of RTL Outcomes}
To evaluate the reliability of our framework, we conducted a manual inspection of the generated RTL across all benchmark suites. Although the Python reference is not 100\% correct, our manual inspection revealed a consistent trend: all generated Verilog code that fails in differential testing is incorrect. As a result, we categorized the generated Verilog code into three groups:

\begin{itemize}
    \item Designs that did not pass the differential testing. 
    \item Designs that passed the differential testing but are functionally incorrect (False Positives). 
    \item Designs that are functionally correct. 
\end{itemize}
    
\subsubsection{Impact of Coverage-Driven Refinement on False Positive Rate (FPR)}

Even with a high-quality reference model and refined inputs from Stage 1, the testbench may still be incomplete, allowing functionally incorrect RTL designs to pass validation. To quantify this risk, we evaluate the false positive rate (FPR), defined as the proportion of designs that pass the testbench validation but are functionally incorrect.

\begin{equation} 
FPR = 1 - \frac{\text{\# designs that are functionally correct}}{\text{\# designs that pass the differential testing}} 
\end{equation}

A lower FPR indicates that the Full Input test suite produced by the Concolic engine is highly reliable, effectively filtering out invalid designs and ensuring that the majority of validated designs maintain high quality.

\begin{figure}
    \centering
    \includegraphics[width=1\linewidth]{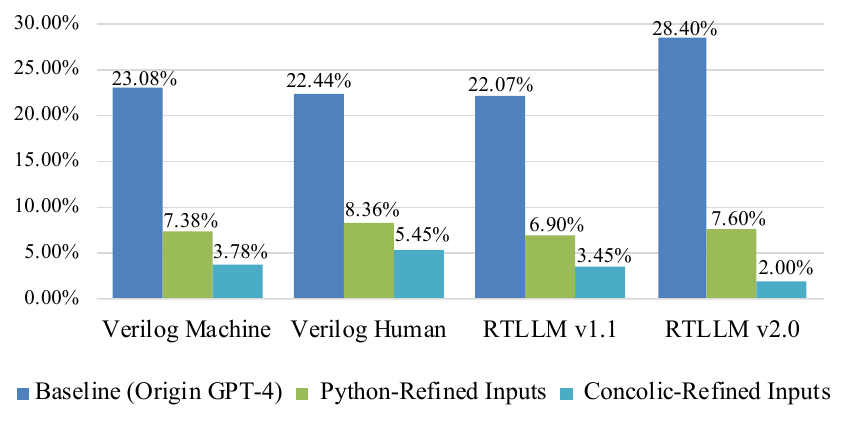}
    \caption{The percentage of designs that passed the differential testing but are functionally incorrect.}
    \label{fig:FPR}
\end{figure}

We evaluated the FPRs across four benchmarks under three distinct configurations: (1) the \textbf{Baseline (Origin GPT-4)}, which utilizes the raw, unrefined testbench from Stage 1; (2) the \textbf{Python-Refined} configuration, where Stage 1 optimizes testbench completeness via feedback; and (3) the \textbf{Concolic-Refined (AutoVeriFix+)} configuration, where Stage 3 introduces the Concolic engine to maximize RTL code coverage. As shown in Fig.~\ref{fig:FPR}, the results confirm that incorporating these successive feedback mechanisms significantly reduces the FPR, effectively filtering out latent functional errors.

Across all benchmarks, the baseline exhibits a significantly high FPR, peaking at 28.40\% on RTLLM v2.0. This indicates that nearly one-third of the designs validated by a naive testbench actually harbor latent functional errors, rendering the baseline output unreliable for direct adoption.  In contrast, integrating Concolic testing successfully intercepts these silent bugs, drastically reducing the FPR to just 2.00\% on RTLLM v2.0 and consistently achieving low rates across other benchmarks.
The step-wise reduction in FPR validates AutoVeriFix+ as a stringent filter, confirms that our Concolic-driven mechanism ensures the validated designs are functionally robust and highly reliable.


\subsection{Evaluation of Concolic Testing}
As presented in Table~\ref{tab:concolic_results}, the integration of Concolic Testing in Stage 3 yields a consistent and significant improvement in Verilog branch coverage across all evaluated benchmarks. While the input vectors derived from the Python reference model (Stage 1) provide a solid baseline—averaging around 88\% coverage—they fail to reach full coverage due to the limitations of random or heuristic-based test generation. Concolic testing in Stage 3 effectively bridges this gap. For instance, on the VerilogEval-human benchmark, coverage surged from 85.22\% to 95.83\%. Furthermore, on the RTLLM v1.1 benchmark, the Concolic engine achieved near-perfect coverage at 99.63\%.

\begin{table}[h!]
\renewcommand{\arraystretch}{1.5}
\centering
\caption{Evaluate the Verilog branch coverage of the input vector generated in Stage 1 using GPT-4 and concolic testing in Stage 3.}
\setlength{\tabcolsep}{1mm}{
\begin{tabular}{l|c|c|c}
\toprule
\multirow{2}{*}{\textbf{Verilog Dataset}} 
  & \textbf{Baseline}  & \textbf{Python-Refined} & \textbf{Concolic-Refined} \\ 
  & \textbf{Coverage (\%)} & \textbf{Coverage (\%)}& \textbf{Coverage (\%)}  \\ \midrule

\textbf{VerilogEval-machine}  & 84.54   & 87.31   &  92.37 \\
\textbf{VerilogEval-human}    & 79.66   & 85.22  &   95.83  \\ 
\textbf{RTLLM v1.1}           & 79.96   & 90.91    &   99.63   \\ 
\textbf{RTLLM v2.0}           & 81.89   & 89.74    &   98.34 \\ 
\bottomrule
\end{tabular}
\label{tab:concolic_results}
} 
\end{table}

This exhaustive coverage directly explains the high reliability and low False Positive Rate discussed in the previous section. By expanding logic coverage to near 100\%, Concolic testing exposes latent defects that would otherwise pass the baseline testbench unnoticed. The drastic drop in FPR on RTLLM v2.0—from 28.40\% to 2.00\%—is directly attributable to the Concolic engine's ability to systematically exercise complex state transitions that the initial testbench missed. The high Verilog coverage rates also provide the mathematical confidence required for the Redundancy Elimination task. For instance, in RTLLM v1.1, the fact that coverage reached 99.63\% implies that the remaining 0.37\% of code is likely unreachable dead code or redundant defensive logic.

\subsection{Evaluation of Redundancy Elimination}

To quantify the physical impact of our redundancy elimination, we synthesized the generated RTL designs using Yosys and compared the average gate counts before and after Stage 3. Table~\ref{tab:optimize_result} presents the results, broken down into Total Cells (including registers) and Combinational Logic Cells. 

For the VerilogEval datasets, which consist of diverse logic problems, we observed a substantial reduction in total gate count: 19.2\% for the machine-generated set and 26.6\% for the human-curated set. This indicates that the initial LLM-generated code often contains superfluous logic branches (e.g., redundant default cases or unreachable else blocks) that inflate the circuit area without contributing to functionality. For the RTLLM v2.0 dataset, which comprises complex and realistic RTL design scenarios, it optimized 10.5\% in total gate count and 4.5\% in combinational logic.

\begin{table}[h!]
\renewcommand{\arraystretch}{1.5}
\centering
\caption{Average Total and Combinational Gate Counts across Benchmarks using Yosys Synthesis.}
\setlength{\tabcolsep}{2.5mm}{
\begin{tabular}{c|c|c|c|c}
\toprule
\multirow{2}{*}{\textbf{Verilog Dataset}} 
& \multicolumn{2}{c|}{\textbf{Before Optimize}} & \multicolumn{2}{c}{\textbf{After Optimize}} \\\cline{2-5}
& \textbf{Total} & \textbf{Comb}  & \textbf{Total} & \textbf{Comb} \\ \midrule
\textbf{VerilogEval-machine}  & 147.84   & 117.93  &  119.42   & 91.65   \\
\textbf{VerilogEval-human}    & 149.46    & 117.68  & 109.69   & 91.18  \\
\textbf{RTLLM v1.1}           & 314.28   & 268.16   &  246.87  &  214.35 \\
\textbf{RTLLM v2.0}           & 360.67   & 310.62   & 322.76   & 296.53      \\
\bottomrule
\end{tabular}
\label{tab:optimize_result}
} 
\end{table}

\subsection{Qualitative Case Study}
To demonstrate the efficacy of Stage 3 in handling complex implementation details, we present a case study using a control logic module shown in Listing~\ref{lst:example2}.  The specification requires the module to increment an internal counter upon receiving a specific input (8'h02) and immediately assert the output out when the counter reaches 1. The initial LLM-generated code (Listing~\ref{lst:example2}) failed this timing requirement. The code checked the value of counter to determine out inside the same clocked block where counter was updated. Since non-blocking assignments do not update the register value until the end of the time step, the logic if (counter == 8'h01) evaluated the old value (0) instead of the new value (1). Consequently, the output out was asserted one clock cycle later than expected.

\begin{figure}[t]
    \centering
    \includegraphics[width=0.95\columnwidth]{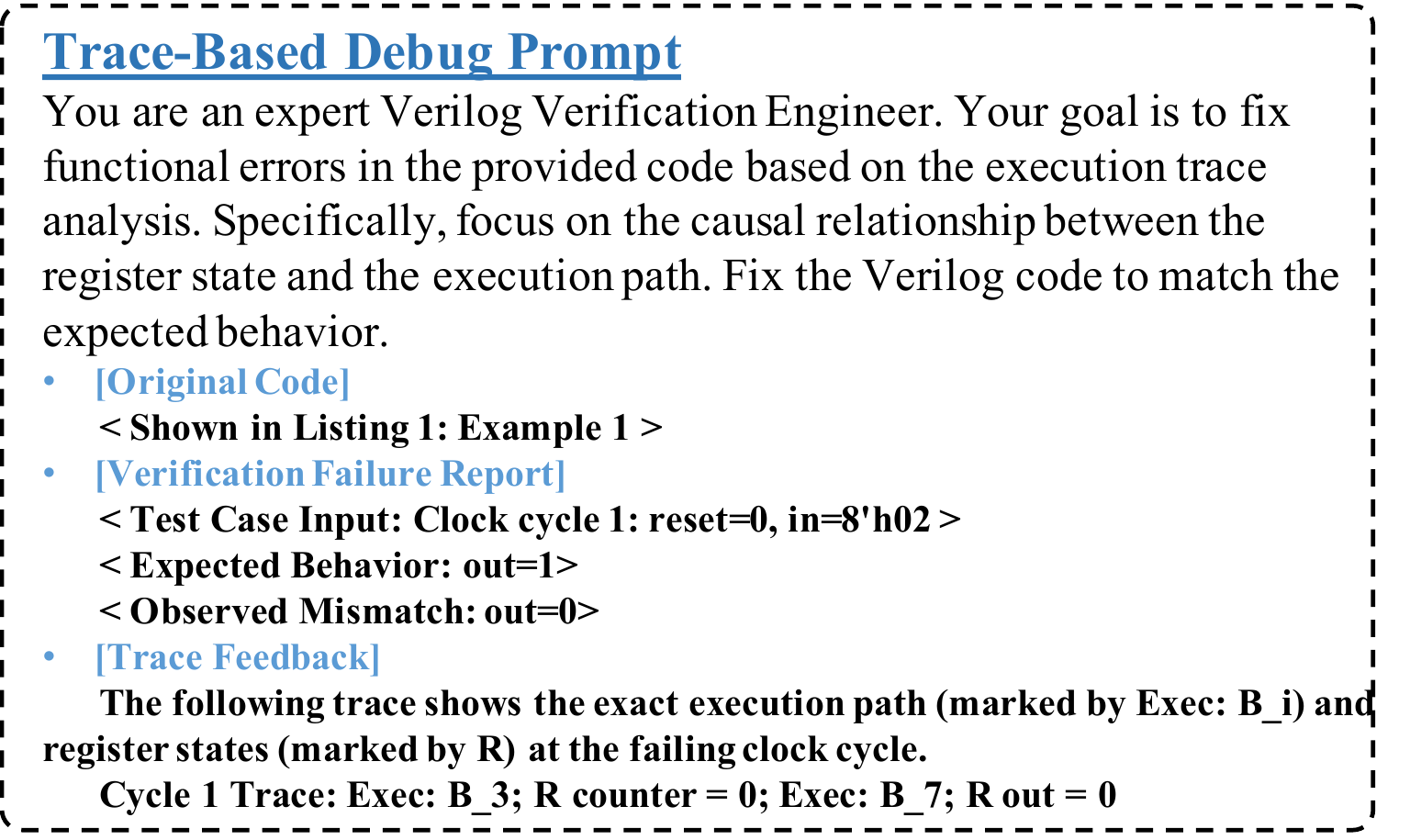}
    \caption{The Example of debugging}
    \label{fig: example}
\end{figure}

If we only feedback the simulation reports a mismatch: “Expected out=1, Got out=0”. With this limited feedback, the LLM often hallucinates, incorrectly modifying the reset logic or polarity, failing to address the root cause. Our framework captures the execution trace and register states during the failing cycles and provides detailed feedback to the LLM, as shown in Fig.~\ref{fig: example}.

\begin{figure}[h]
\begin{lstlisting}[caption={Optimized Verilog}, label={lst:example2}]
module top (
    input  wire       clock,
    input  wire       reset,
    input  wire [7:0] in,
    output reg        out
);
    reg [7:0] counter;
    reg [7:0] next_counter; // [FIX] Introduce intermediate state
    // 1. Combinational Logic for Next State
    always @(*) begin
        case (in)
            8'h00:   next_counter = counter - 1;
            8'h02:   next_counter = counter + 1;
            8'hFF:   next_counter = 8'h00;
            // [OPTIMIZATION] 'default' case removed based on UNSAT proof
        endcase
    end
    // 2. Sequential Logic Update
    always @(posedge clock) begin
        if (reset) begin
            counter <= 8'h00;
            out <= 1'b1;
        end
        else begin
            counter <= next_counter;
            // [FIX] Check 'next_counter' (the new value) instead of 'counter'.
            // This ensures 'out' reflects the state AFTER the update.
            if (next_counter == 8'h01)
                out <= 1'b1; 
            else
                out <= 1'b0;
        end
    end
endmodule
\end{lstlisting}
\end{figure}

The optimized code produced after receiving Stage 3 feedback demonstrates significant improvements.  Firstly, guided by the trace feedback, the LLM refactored the design to separate combinational logic (next\_counter) from sequential logic. This explicitly resolves the timing race by checking next\_counter, ensuring the out signal aligns with the updated state. With the coverage report, the LLM successfully removed the redundant default branch, resulting in a cleaner and more area-efficient design.

\section{Conclusion}
\label{subsec:conclusion}

This paper presents a three-stage framework for RTL code generation that harnesses the strengths of LLMs in Python code generation to create a Python-based reference model. In the first stage, the framework establishes Python reference models as functional standards, offering high-level implementations that guide the creation of testbenches. A feedback mechanism is incorporated, enabling the LLM to enhance testbench coverage by analyzing coverage metrics. In the second stage, the LLM iteratively addresses syntax errors with compiler feedback and corrects functional errors by identifying inconsistencies between the Verilog implementation and the testbench. These iterative improvements ensure that the generated Verilog RTL maintains a high level of functional correctness. Experimental results demonstrate that our approach outperforms both existing general-purpose LLMs and hardware-specific models in producing accurate Verilog code.

\bibliographystyle{IEEEtran}
\bibliography{references}




\end{document}